\documentclass{iau} 
\usepackage{graphicx}

\title[Galaxies Blurred by the Planck Scale] 
{Limits to Seeing High-Redshift Galaxies \\Due to Planck-Scale-Induced Blurring}

\author{Eric Steinbring}   

\affiliation{National Research Council Canada, Herzberg Astronomy and Astrophysics,\\
Victoria, British Columbia, Canada V9E 2E7
\\email: {\tt Eric.Steinbring@nrc-cnrc.gc.ca}}

\pubyear{2015}
\volume{319}  
\jname{Galaxies at High Reshift and Their Evolution over Cosmic Time}
\editors{S. Kaviraj, ed.}
\begin{document}

\maketitle




\vspace{1em}

If spacetime is ``foamy" travel along a lightpath must be subject to continual, random distance fluctuations $\pm \delta l$ proportional to Planck length $l_{\rm P} \sim {10}^{-35}~{\rm m}$ (Lieu \& Hillman 2003). Although each ``kick" by itself is tiny, these may accumulate.  Accounting for redshifted (bluer) emitted photons, over a cosmological distance $L = (1+z)L_{\rm C}$ for co-moving distance $L_{\rm C}$, the resultant phase perturbations $\Delta \phi = 2\pi \delta l/\lambda$ at observed wavelength $\lambda$ could grow independently of telescope diameter $D$ to a maximum of $\Delta\phi_{\rm max}=(1+z)\Delta\phi_0$ (Steinbring 2007) where $\Delta\phi_0=2\pi a_0 (l_{\rm P}^{\alpha}/\lambda)L^{1 - \alpha}$ follows Ng et al. (2003). Here $a_0\sim 1$ and $\alpha$ specifies the quantum-gravity model: $1/2$ implies a random walk and $2/3$ is consistent with the holographic principle; a vanishingly small $\Delta\phi_{\rm P}=\Delta\phi_{\rm max}/[(1 + z) a_0 (L/l_{\rm P})^{1 - \alpha}]=2\pi l_{\rm P}/\lambda$ is approached when $\alpha=1$.

The highest-resolution Nyquist-sampled images of $z\approx6$ active galactic nucleii (AGN) ever obtained are suggestive although do not show clear evidence of $\Delta \phi_{\rm max}$ for $\alpha=2/3$ (Steinbring 2007; Tamburini et al. 2011). That level of blurring, however, would imply an image full-width at half maximum (FWHM) just at the diffraction limit of the 2.4-m diameter {\it Hubble Space Telescope} ({\it HST}). It might be that $\alpha$ is larger or that these phase errors become invisible as they approach the wavelength of observed light (e.g. Perlman et al. 2015). But it can be shown by taking a linear superposition of all phase-error amplitudes ${\Delta \phi}~\sigma (\Delta \phi) = 1-A \log({{\Delta \phi}/{\Delta \phi_{\rm P}}})$ that $(1/A)\int \Delta \phi ~\sigma (\Delta \phi) ~{\rm d}{\Delta \phi} = (1 + z) \Delta \phi_0$ is recovered for $A = 1/\log{[(1 + z)(L/l_{\rm P})^{1 - \alpha}]}$. If correct, any pointlike source viewed by a diffraction-limited telescope for which $\Delta\phi_{\rm max}$ is visible must be inflated to a lesser mean width of
$$\Phi = 1.22 {\lambda\over{D}} + \int \Delta \phi ~\sigma (\Delta \phi) ~{\rm d}{\Delta \phi} \approx 1.22 {\lambda\over{D}} + 2 \pi {l_{\rm P}\over{\lambda}} A e^{1/A}.$$ 
That is consistent with the {\it HST} results as well as {\it Fermi} observations of $z<4$ gamma-ray bursts (GRBs; Steinbring 2015). Unfortunately, those and current ground-based telescopes employing adaptive optics (AO) are unable to reliably resolve structures much below 0.3 kpc wide at $z=1$ to 4 (assuming a cosmology with $\Omega_\Lambda=0.7$, $\Omega_{\rm M}=0.3$, and $H_0=70~{\rm km}~{\rm s}^{-1}~{\rm Mpc}^{-1}$). Future optical/near-infrared AO and space telescopes viewing out to $z=6$ to 8 will do better. In a long exposure, the minimal angular extent of known pointlike galaxy structures would be noticeably enlarged - a FWHM expanded by more than a few percent - and possibly also embedded within an extra halo of scattered ``fuzz." For example, with $z=6.0$, $\alpha=0.667$, and $D=6.5~{\rm m}$ no AGN or GRB, despite being physically much smaller, would be found with an apparent size less than about 0.1 kpc in the restframe, 0.025 arcsec across, or consistently 8\% larger than diffraction at $0.6~\mu{\rm m}$, the shortest wavelength to be visible with the {\it James Webb Space Telescope}.

\end{document}